\documentclass[aps,showpacs,twocolumn,superscriptaddress,amssymb]{revtex4}
\usepackage{graphicx,epsfig,latexsym,overpic,amssymb,color}
\preprint{\today}
\begin{document}
\title{Density distributions of superheavy nuclei}
\author{J.C. Pei}
\affiliation{School of Physics, Peking University, Beijing 100871, China}
\author{F.R. Xu}
\email{frxu@pku.edu.cn}
\affiliation{School of Physics, Peking University, Beijing 100871, China}
\affiliation{Institute of Theoretical Physics, Chinese Academy of Sciences,
Beijing 100080, China}
\affiliation{Center of Theoretical Nuclear Physics. National Laboratory of
Heavy Ion Collisions, Lanzhou 730000, China}
\author{P.D. Stevenson}
\affiliation{Department of Physics, University of Surrey,
Guildford GU2 7XH, United Kingdom}

\begin{abstract}
We employed the Skyrme-Hartree-Fock model to investigate the
density distributions and their dependence on nuclear shapes and
isospins in the superheavy mass region. Different Skyrme forces
were used for the calculations with a special comparison to the
experimental data in $^{208}$Pb. The ground-state deformations,
nuclear radii, neutron skin thicknesses and $\alpha$-decay
energies were also calculated. Density distributions were
discussed with the calculations of single-particle wavefunctions
and shell fillings. Calculations show that deformations have
considerable effects on the density distributions, with a detailed
discussion on the $^{292}$120 nucleus. Earlier predictions of remarkably
low central density are not supported when deformation is allowed for.

\end{abstract}
\maketitle
\section{introduction}
Proton and neutron density distributions, the associated
root-mean-square (rms) radii $R_{\rm p}$ and $R_{\rm n}$,
and the neutron skin thickness $\Delta R=R_{\rm n}-R_{\rm p}$
provide fundamental information on
nuclear structure. For example, halo nuclei are charactered with
long tails in density distributions. The density is a direct probe
of the size of an atomic nucleus, and plays an important role in
the cross sections of nuclear reactions. While charge densities
can be measured from the elastic scattering of electrons, neutron
densities are largely unknown. The parity-violating electron
scattering has been suggested to measure neutron densities
\cite{Donnelly89}. With the experimental method, the neutron
densities of more nuclei can be expected to be measured.
Theoretically, the calculated neutron skin thickness can be
model dependent \cite{Horow01}.

Recently, Horowitz {\it et al.} \cite{Horowitz01} studied the
relationship between the neutron skin of the spherical double magic
nucleus $^{208}$Pb and the properties of neutron-star crusts,
showing the importance of the knowledge of nucleon densities in
understanding the equation of state of neutron-rich matter and
therefore the properties of neutron stars. The heavy nucleus
$^{208}$Pb has been measured to have a neutron skin thickness of about
0.15 fm \cite{trzcinska}. With increasing mass number,
the neutron excess becomes larger in general and it's natural to
think that superheavy nuclei provide the largest neutron excesses.
The heaviest nuclei also have large proton numbers and thus large
Coulomb repulsive forces that push the protons to larger radii and
therefore change density distributions. Novel density
distributions were predicted in extraordinary $A>400$ nuclei that
have bubbles and showed coupling effects between density
distributions and shell
structures \cite{Dietrich98,Decharge99,Yu00}. The nucleon density
in a bubble is reduced to be zero. Semi-bubble nuclei were also
suggested with the considerable reduction of central densities for
the $Z\geq 120$ nuclei \cite{Decharge99} located around the
center of the predicted island of stability of superheavy nuclei.
Bender {\it et al.} have investigated the density distributions of
superheavy nuclei with the restriction of spherical shapes
\cite{bender99}. Recent progress in experiments are motivating the
structure study of superheavy
nuclei \cite{Hofmann00,Oganessian99}. Many theoretical works have
investigated the properties of superheavy nuclei
\cite{cwiok96,cwiok,xu,smolanczuk97,nazarewicz,bender00r,ren01,ren,meng,wu},
such as shell structure, $\alpha$ decay and spontaneous fission.
Experiments have also provided the structure information of superheavy
nuclei by the in-beam study of spectroscopy \cite{Reiter99,Herzberg01,Butler02,
Herzberg04}. In the present work, we investigate the density distributions of
superheavy nuclei and related structure properties,
with deformation effects taken into account.

\section{calculations }

The deformed Skyrme-Hartree-Fock model (SHF) \cite{blum} was used
in the present investigation. Pairing correlations are treated in
the BCS scheme using a $\delta$-pairing force,
$V_{\rm pair}=-V_{\rm q}\delta(\vec{r}_1-\vec{r}_2)$ \cite{krieger,bender00}.
The pairing strength $V_{\rm q}$ (q=p, n for the protons and neutrons,
respectively) has been parameterized throughout the chart of
nuclei \cite{bender00}, but the actual values are dependent on
Skyrme forces chosen. The detailed values of the pairing
strengthes can be found in Ref.~\cite{bender99}.

Calculations are performed in coordinate space with axially
symmetric shape. The density distribution of protons or neutrons
is given in the two-dimensional form as follows.
\begin{equation}
 \rho_{}(z,r)=\sum\limits_{k}
2v_{k}^{2}(|\psi_{k}^{+}(z,r)|^{2}+|\psi_{k}^{-}(z,r)|^{2})
\end{equation}
where, $\psi_{k}^{+}$ and $\psi_{k}^{-}$ are the components of the
wavefunctions with intrinsic spin s$_{\rm z}$=$+\frac{\hbar}{2}$ and
$-\frac{\hbar}{2}$, respectively, and $v_k^2$ is the pairing
occupation probability of the $k$-th orbit. The ground states of
most superheavy nuclei are expected to have axially symmetric or
spherical shape \cite{smolanczuk97}. In this paper, we consider
the most important axially symmetric deformations, $\beta_2$ and
$\beta_4$.

In the present work, we investigated the densities and related
structure problems of even-even superheavy nuclei with
Z=104$-$120. In the SHF calculations, results are in general parameter
dependent. For example, the SkI3 force predicts $^{292}$120 for
the next magic nucleus beyond $^{208}$Pb, while SLy7 and SkI4
predict $^{310}$126 and $^{298}$114 for the magic nucleus,
respectively \cite{cwiok96,bender99}. To make comparison, we
used the different sets of parameters SLy4,
SLy7 \cite{chabanat}, SkI3 and SkI4 \cite{reinhard}. These
sets of parameters have been developed recently with good isospin
properties, and we note that they have been recommended by Rikovska
Stone {\it et al.} for their ability to describe realistic neutron
stars and the properties of asymmetric nuclear matter \cite{Rik03}.
In the superheavy mass region, Skyrme parameter sets can reproduce
experimental binding energies within a few MeV \cite{burvenich}
and $\alpha$-decay energies within a few hundred keV \cite{cwiok}.
Table I lists the properties of the ground states calculated with SkI4 for
experimentally known even-even superheavy nuclei and the predicted
magic nucleus $^{298}$114. The calculated $\alpha$-decay energies
agree with experimental data within a few hundred keV (The largest
difference with data is 640 keV in $^{266}$Hs). Experimental binding energies
can be reproduced within $\approx 4$ MeV for the nuclei listed in Table I
with the SkI4 force. The obtained neutron-skin thicknesses are
smaller than the calculations by the relativistic mean-field (RMF)
\cite{ren}. (It was pointed out that RMF calculations
overestimate neutron-skin thicknesses \cite{Furnstahl02}.)

\begin{figure}[t]
\includegraphics[width=8cm]{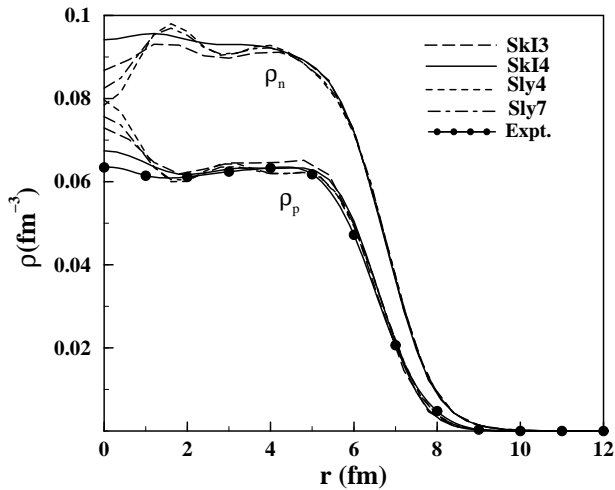}
\label{Fig.1} \caption{The calculated proton ($\rho_{\rm p}$) and
neutron ($\rho_{\rm n}$) density distributions of the spherical
nucleus $^{208}$Pb, compared with the experiment for the charge
density \cite{pb208}.}
\end{figure}

\begin{figure}
\includegraphics[width=8cm]{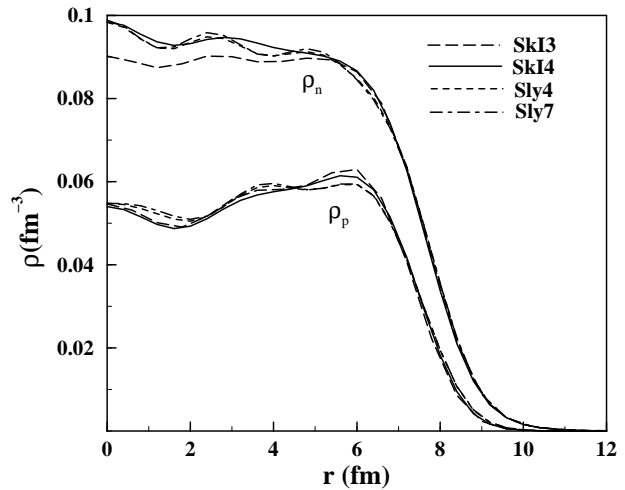}
\label{Fig.2}\caption{The calculated proton and neutron densities
for spherical nucleus $^{298}$114. }
\end{figure}

\begin{table}[b]
\caption{The SkI4 calculations for the experimentally synthesized
even-even superheavy nuclei. The calculated $\alpha$-decay
energies ($Q_{\alpha}^{\rm cal.}$) are compared with experimental
values ($Q_{\alpha}^{\rm expt.}$). The experimental data are taken
from \cite{audi}. The calculation for the predicted magic nucleus
$^{298}$114$_{184}$ is also listed.}
\begin{ruledtabular}
\begin{tabular}{cccccccc}
nuclei&$\beta_2$&$\beta_4$&$R_{\rm n}$&$R_{\rm p}$&$\Delta R$
&$Q_{\alpha}^{\rm cal.}$&$Q_{\alpha}^{\rm expt.}$\\
&&&(fm)&(fm)&(fm)&MeV&MeV\\
\hline
$^{254}$Rf&0.29&0.06&6.05&5.96&0.09&9.35&9.38\\
$^{256}$Rf&0.29&0.04&6.07&5.97&0.10&9.47&8.96\\
$^{258}$Rf&0.28&0.03&6.09&5.98&0.11&9.07&9.25\\
$^{258}$Sg&0.28&0.03&6.08&5.99&0.09&10.04&9.70\\
$^{260}$Sg&0.28&0.01&6.10&6.00&0.10&9.65&9.93\\
$^{266}$Sg&0.26&-0.03&6.15&6.03&0.12&8.28&8.88\\
$^{264}$Hs&0.27&-0.02&6.12&6.03&0.09&10.24&10.59\\
$^{266}$Hs&0.26&-0.03&6.14&6.04&0.10&9.70&10.34\\
$^{270}$Hs&0.25&-0.06&6.18&6.06&0.12&8.94&9.30\\
$^{270}$110&0.25&-0.05&6.16&6.07&0.09&11.32&11.20\\
$^{284}$112&0.15&-0.07&6.25&6.12&0.13&9.65&9.30\\
$^{288}$114&0.13&-0.08&6.27&6.14&0.13&10.36&9.97\\
$^{292}$116&0.06&-0.02&6.27&6.15&0.12&10.64&10.71\\
$^{298}$114&0.0&0.0&6.33&6.16&0.16&7.4&\\
\end{tabular}
\end{ruledtabular}
\end{table}

\begin{figure}[t]
\includegraphics[width=8cm]{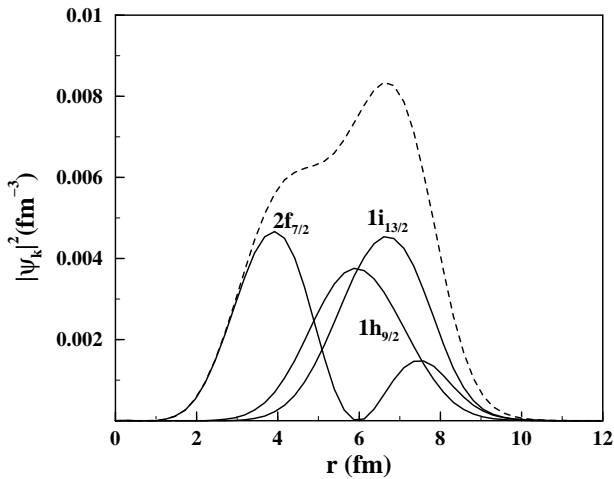}
\label{Fig.3}\caption{The SkI4 calculated $|\psi _k|^2$
distributions of the proton 1i$_{13/2}$, 1h$_{9/2}$ and 2f$_{7/2}$
orbits in the case of spherical $^{298}$114 nucleus. The dashed
line stands for the summed contribution of the orbits. }
\end{figure}

\subsection{Density distributions of spherical nuclei}

In order to test potential parameters in the calculations of
densities, we calculated the density distributions of the
spherical doubly magic nucleus $^{208}$Pb using the different sets
of parameters. For $^{208}$Pb, the charge distribution has been
measured by electron scattering \cite{pb208}. Fig.1 shows the
calculated density distributions with comparison with the
experimental charge density. It can be seen that the proton
density given by the SkI4 force is closest to the experimental
measurement. The rms radii of $^{208}$Pb are calculated with the
SkI4 force to be 5.43 fm for the protons and 5.61 fm for the
neutrons, leading to a neutron-skin thickness of 0.18 fm which is
slightly larger than the values of 0.16 fm given by the SLy4 and
SLy7 forces. These results agree with the 0.15$\pm$0.02 fm from
the recent antiprotonic atom experiment \cite{trzcinska}. The SkI3
force gives a larger neutron skin thickness of 0.23 fm compared to
the other three forces, which may be due to the similar behavior
of SkI3 force to the RMF model \cite{bender99}. It needs to be
pointed out that nuclear ground-state correlations (see, e.g.,
\cite{Strayer,Dang01}) can have visible effects on nuclear
properties, such as energies and densities. The oscillations
observed in the calculated density distributions in Fig.1 could be
reduced when the correlation is taken into account. Such
correlations, which go beyond the mean-field approximation, are
not included in the present work.

\begin{figure}[t]
\includegraphics[width=8cm]{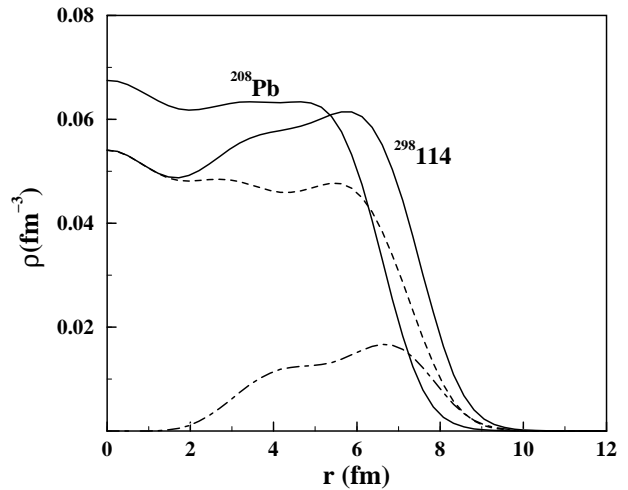}
\label{Fig.4}\caption{The SkI4 calculated proton densities of
$^{298}$114 and $^{208}$Pb. In the $^{298}$114 nucleus, the dashed
line presents the contribution from the orbits below the Z=82
closed shell, while the dot-dashed line presents the contribution
from the high-$j$ orbits of 1i$_{13/2}$, 1h$_{9/2}$ and 2f$_{7/2}$
that construct the next closed shell. }
\end{figure}

\begin{figure}[b]
\includegraphics[width=8cm]{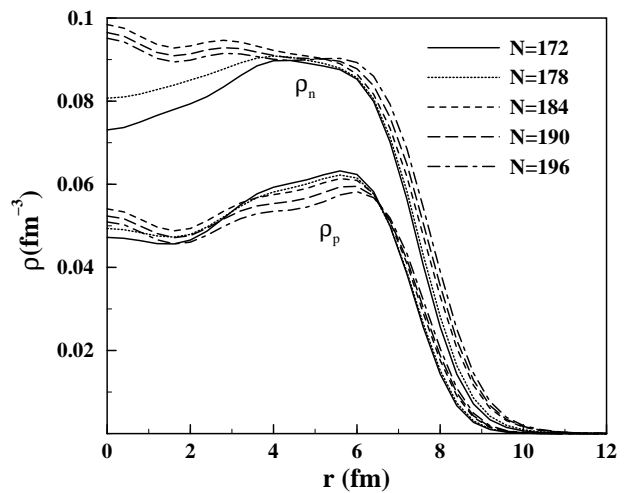}
\label{Fig.5}\caption{The spherical density profiles of the nuclei
with Z=114 and N=172$-$196 calculated with the SkI4 force.}
\end{figure}

Fig.2 shows the calculated densities of the spherical nucleus
$^{298}$114. This nucleus was predicted to be the next doubly closed shell
nucleus by a Macro-microscopic
model \cite{smolanczuk97} and SHF with SkI4
force \cite{bender99}. For $^{298}$114, the calculations with
SkI4, SLy7 and SLy4 give similar density distributions. It can be
seen that the charge density distribution of $^{298}$114 has a
central depression. The central density depression has been
predicted to exist widely in the spherical superheavy nuclei
\cite{bender99}. Fig.3 displays the calculated square
wavefunctions of the proton 1i$_{13/2}$, 1h$_{9/2}$ and 2f$_{7/2}$
orbits that locate the 82$<$Z$\leqslant$114 closed shell. It can
be seen that the high-$j$ orbits have density contributions in the
nuclear surface region. This is consistent with the classical
picture in which orbits with large angular momentum locate at
surface. In Fig.4, we show the proton densities of $^{208}$Pb and
$^{298}$114 for comparison. The proton density of $^{298}$114 is
decomposed into two parts: i)the contribution from proton orbits
below the Z=82 closed shell; and ii) from the orbits in the next
closed shell with 82$<$Z$\leq$114. The contribution from the
orbits below Z=82 has a similar behavior to the charge distribution
of $^{208}$Pb, without central depression. The 82$<$Z$\leq$114
orbits (that have high-$j$ values) have contributions in the
surface region of the nucleus, leading to a central depression in
the proton density of $^{298}$114.

For the neutrons at spherical case, the high-$j$ orbits of
2g$_{9/2}$, 1i$_{11/2}$, 1j$_{15/2}$ and 2g$_{7/2}$ occur in the
region of N=126$-$172. The low-$j$ orbits of 4s$_{1/2}$ and
3d$_{3/2}$ are at N=178$-$184. Fig.5 shows the density
distributions for N=172$-$196 and Z=114 with assumed spherical
shape. Indeed, these nuclei were predicted to be nearly spherical
in their ground states \cite{smolanczuk97}. It can be seen that
proton densities have central depressions and neutron densities
become centrally depressed for N$\leq$178.

\begin{figure*}[bhpt]
\vspace{-35pt} \hspace{-124.pt}
\includegraphics[width=20cm,height=28cm]{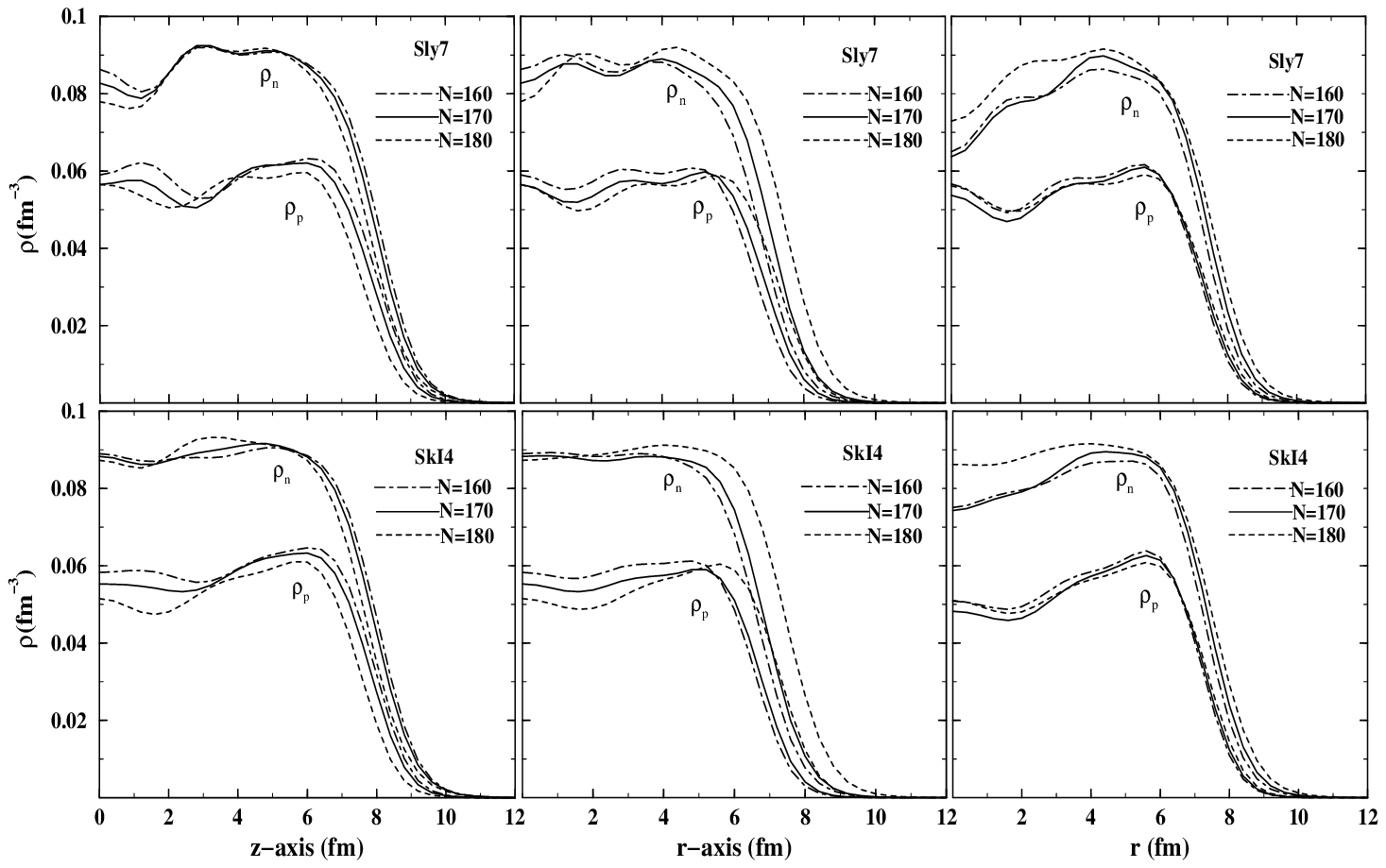}
\hspace{-100.pt} \vspace{-495pt} \label{Fig.6}\caption{ Calculated
density distributions of nuclei of the deformed
$^{270,280,290}$110. The left and middle columns show the
distribution profiles of $\rho(z,0)$ along the $z$-axis and
$\rho(0,r)$ along the $r$-axis, respectively. For comparison,
distributions at assumed spherical shape are also displayed in the
right column.}
\end{figure*}

\begin{figure}[b]
\includegraphics[width=7cm,height=7cm]{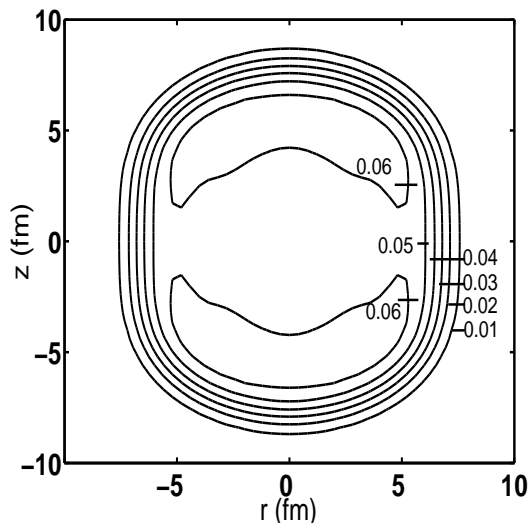}
\label{Fig.7}\caption{Two-dimensional proton density contours of
the nucleus $^{280}$110, calculated with the SkI4 force. The
numbers on contours are the values of densities in fm$^{-3}$.}
\end{figure}

\begin{figure}
\vspace{-36pt}
%\begin{minipage}[b]{0.5 \textwidth}
\hspace{-470pt}
\includegraphics[width=20cm,height=28cm]{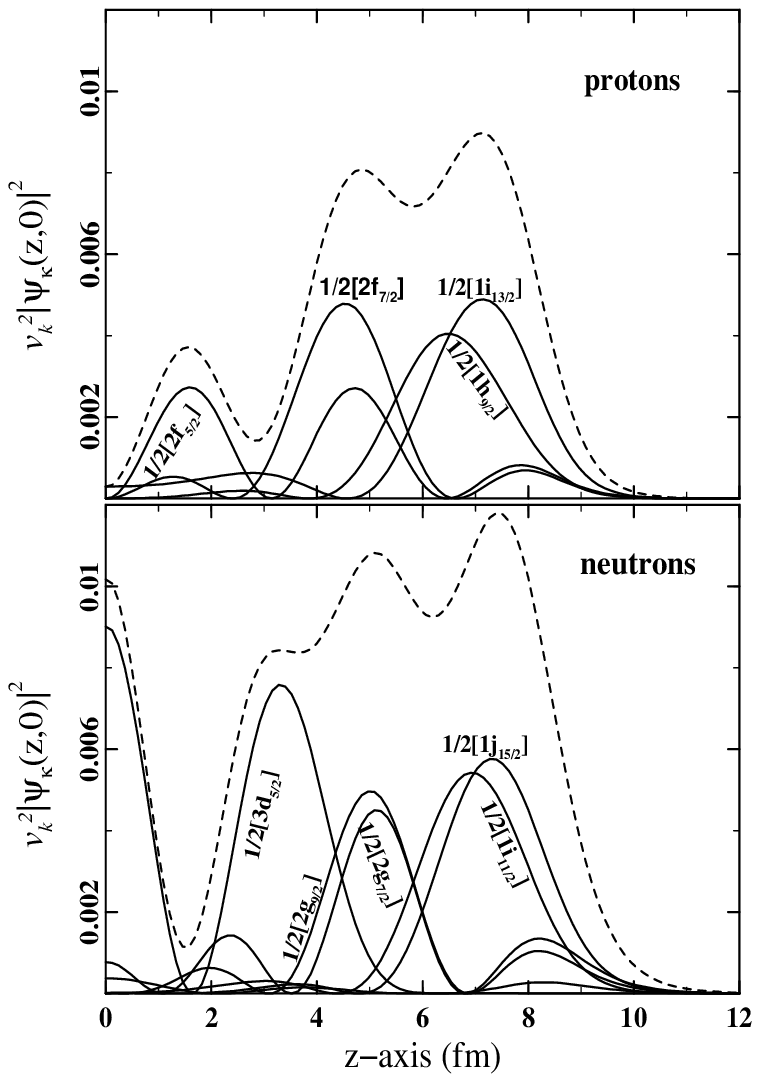}
\hspace{-450pt}
 \vspace{-442pt}
%\end{minipage}
\label{Fig.8}\caption{ The SkI4-calculated contributions (in
$z$-axis) of the $j_z$=1/2 orbits in the $^{280}$110 case, with
the modifications of pairing occupation probabilities. The dashed
line represents the summed contributions of the orbits.}
\end{figure}

\subsection{Densities of deformed superheavy nuclei}

The shell fillings of nucleons are sensitive to the deformations
of nuclei. Therefore, density distributions should be expected to
be shape dependent. Most superheavy nuclei known experimentally
are believed to have deformed shapes. Fig.6 displays the
calculated density distributions for the Z=110 isotopes with
N=160, 170 and 180 nuclei. The equilibrium deformations are
determined by minimizing calculated energies. The deformations
determined with the SkI4 force are $\beta_2$=0.25, 0.19 and 0.05
for $^{270,280,290}$110, respectively. To check the possible
parameter dependence, we also used the SLy7 force to calculate the
densities, shown in Fig.6. The deformations determined with SLy7
are $\beta_2$=0.25, 0.17 and 0.05 for $^{270,280,290}$110,
respectively. In order to see deformation effects, we calculated
the densities assuming spherical shape. It can be seen for
$^{270,280}$110 that the densities become more centrally
depressed in spherical cases. We also see some difference between
the densities along the $z$- and $r$-axes. Fig.7 shows the
two-dimensional proton density distribution for the $^{280}$110
nucleus. The density has two humps in the $z$-axis. The
double-hump distribution has been suggested experimentally, e.g.,
in the deformed $^{166}$Er and $^{176}$Yb \cite{cooper}. Our
calculations with SkI4 and SLy7 show that such double-hump
phenomenon is relatively pronounced for nuclei around $^{280}$110. The
macro-microscopic calculations show that the even-even nuclei
around N=170 are particularly unstable against spontaneous
fission ({see Fig.9 in \cite{smolanczuk97}). This could be
related to the double-hump distributions in these deformed nuclei.

Fig.8 shows the distributions of the square wavefunctions of the
$j_z=1/2$ orbits in $^{280}$110, modified with pairing occupation
probabilities (see Eq.(1)). These orbits are above the Z=82 and
N=126 shells for the protons and neutrons, respectively. Shown as
in Fig.8, the high-$j$ low-$j_z$ orbits have important
contributions to densities in the surface region of the $z$-axis.
High-$j$ low-$j_z$ orbits have strong prolate-driving effect.
Hence, sufficient number of high-$j$ low-$j_z$ orbits occupied can result in
prolate shapes and double-hump densities. In $^{166}$Er and
$^{176}$Yb that were suggested experimentally to have double-hump
densities, the low-$j_z$ orbits of the proton 1h$_{11/2}$,
1g$_{7/2}$ subshells are occupied.

Density distributions given by different Skyrme forces can differ
as shown in Fig.6. The SLy7 predicts larger central
depression in neutron densities and less central depression in
proton densities than the SkI4 force. This difference also occurs
in the calculation of $^{208}$Pb, see Fig.1. For different
parameters, the opposite behavior of proton and neutron
distributions reduce the difference in the total (proton+neutron)
nuclear density distributions. The origin of the opposite behavior
would be due to the self-consistent coupling between protons
and neutrons in the SHF model, to approach the nuclear density
saturation \cite{richter}. The SkI4 force is better in
reproducing the density of $^{208}$Pb, compared to other Skyrme
forces used in the present investigation. However, the SkI4 force
was pointed out to have larger spin-orbit splittings in the
calculations of single-particle level schemes for the superheavy
region \cite{bender99}.

\begin{figure}[b]
\begin{minipage}[b]{0.5 \textwidth}
\includegraphics[width=7.9cm,height=7cm]{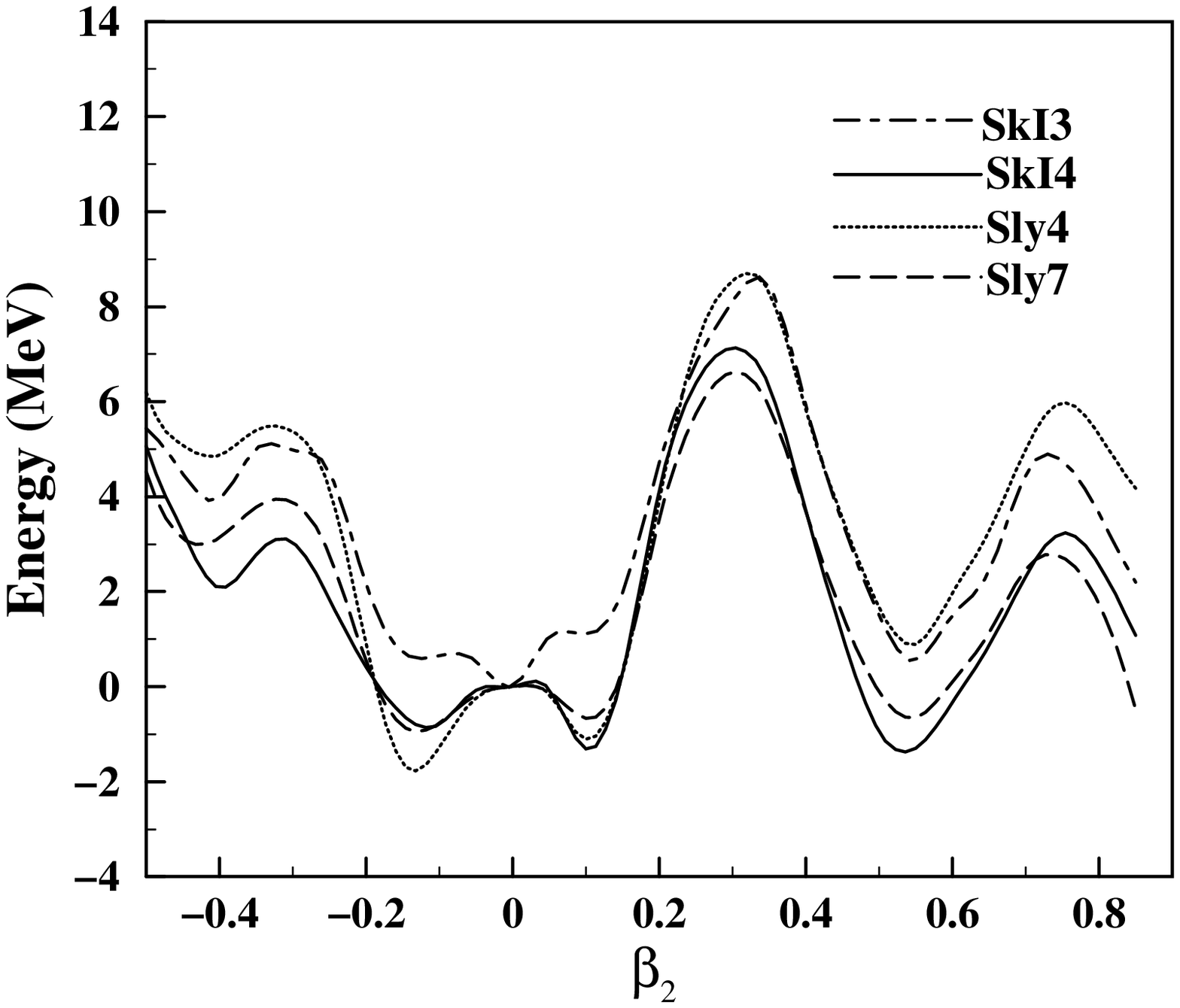}
\end{minipage}
 \label{Fig.9}\caption{Calculated energy curves of
$^{292}$120 with different Skyrme forces, as a function of
quadrupole deformation.}
\end{figure}

\begin{figure*}
\vspace{-85.8pt} \hspace{-240pt}
\includegraphics[width=18cm,height=27cm]{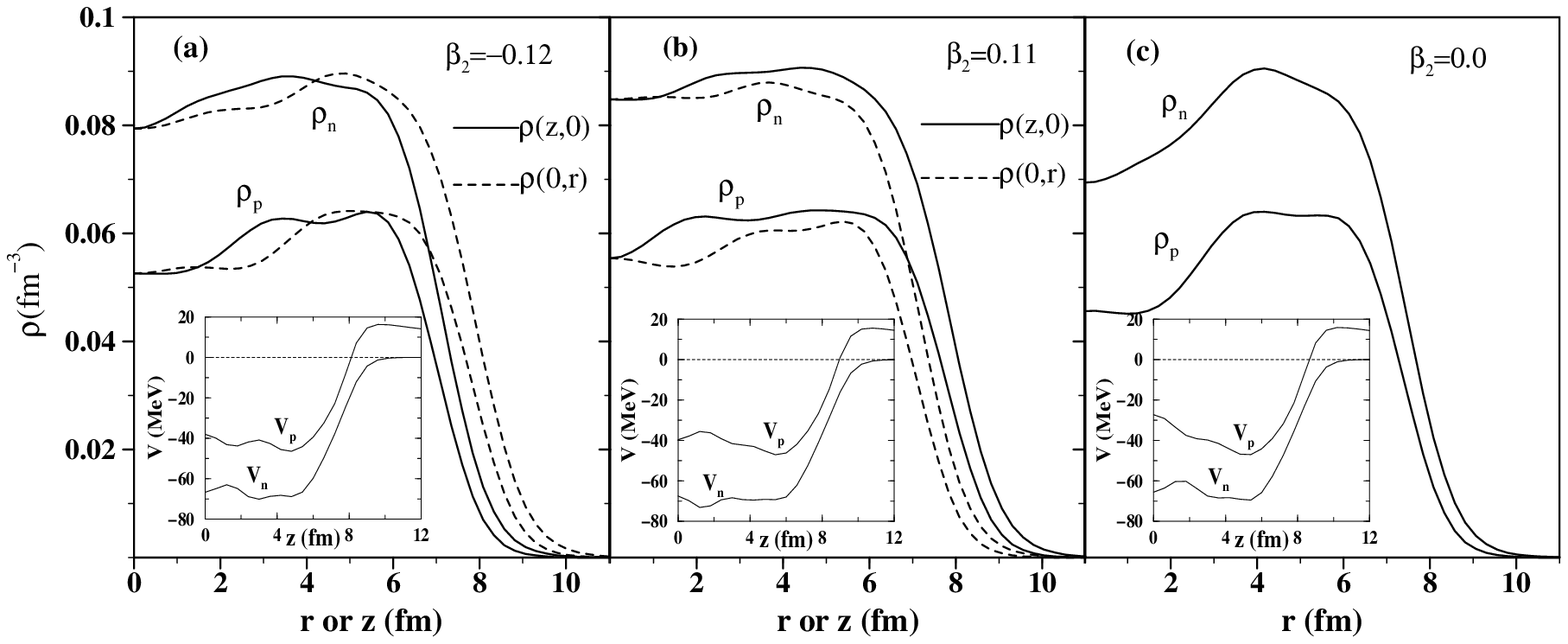}
\hspace{-220pt} \vspace{-455.8pt} \label{Fig.10}\caption{The SkI4
calculated density distributions $\rho(z,0)$ and $\rho(0,r)$ in
$^{292}$120 at $\beta_2=-0.12$ (a), 0.11 (b) and 0.0 (c),
corresponding to the energy minima shown in Fig.9 and the
spherical shape. The inset figures show the single-particle
potentials in the $z$-axis.}
\end{figure*}

\subsection{The $^{292}$120 nucleus}

$^{292}$120 is an interesting nucleus that was predicted to be a
doubly magic \cite{bender99} and spherical semi-bubble
nucleus \cite{Decharge99}. We calculated the energy curve with
the SLy4, SLy7, SkI3 and SkI4 parameters, see Fig.9. The results
of SLy4, SLy7 and SkI4 are close to the Hartree-Fock calculations
with Gogny force \cite{Decharge99}. The SkI3 force gives a shallow
spherical minimum. The SkI4 calculation predicts three minima at
$\beta_2=$0.11, $-0.12$ and 0.52 (Superdeformations in superheavy
nuclei have been discussed by Ren {\it et al.} \cite{ren01,ren}).
It needs to be mentioned that the present calculations are restricted
to axially symmetric shapes without considering the possibility of triaxiality.
The inclusion of the triaxial degree of freedom could alter the shallow
minima. Fig.10 shows the SkI4 calculated density distributions at the
prolate and oblate shapes. For comparison, the densities at the
spherical shape are also displayed. It can be seen that
significant central depressions or central semi-bubble appear in
the spherical case. However, the situation is considerably altered
even with a small shape change. Only weak central depressions are
seen at the small deformations, see Fig.10. To have a further
understanding, we calculated the corresponding single-particle
potentials in the $z$-axis, shown inside Fig.10. It can be seen
that the proton potential has a considerable change with changing
the deformation. The spherical proton potential has a significant
hump at the center of the nucleus. This implies that the Coulomb
energy can be considerably reduced by forming the center
semi-bubble at the spherical shape. The Coulomb energy can also be
reduced by generating the deformation of the nucleus. In the
reduction of the total energy of the nucleus, there is competition
between forming the central semi-bubble and generating the
deformation. The deformation can affect the shell structure and
then density distributions, and vice versa. In a self-consistent
model, such as the SHF approach, densities are fed back into the
potential, which amplifies the coupling between deformations and
densities.

\section{summary}

In summary, the density distributions of superheavy nuclei have
been investigated with the Skyrme-Hartree-Fock model. To test the
model and parameters, the $\alpha$-decay energies of even-even
superheavy nuclei and the charge density of $^{208}$Pb are
calculated and compared to existing experimental data. For
axially-symmetrically deformed nuclei, the density distributions
in the symmetric $z$- and the vertical $r$-axes or the
two-dimension distributions were calculated. The distribution in
different directions can be different. The high-$j$ low-$j_z$
orbits have important contributions to the densities at nuclear
surface in the $z$-axis, while high-$j$ high-$j_z$ orbits have
important contributions at surfaces in the $r$-axis. The
deformation effect was found to be significant in the calculation
of the density distribution in the $^{292}$120 nucleus. Only a weak
central depression was seen in the deformed case for $^{292}$120,
compared to the predicted semi-bubble at the spherical shape.

\acknowledgments
We thank Prof. P.M. Walker for his valuable
comments, and Prof. P.-G. Reinhard for the computer code. This
work was supported by the Chinese Major State Basic Research
Development Program No. G2000077400, the Natural Science
Foundation of China (Grants No. 10175002 and No. 10475002), the
Doctoral Foundation of Chinese Ministry of Education
(20030001088), the U.K. Royal Society, and the U.K. Science and
Engineering Research Council.

\end{document}